\documentclass[aps,pre,reprint,superscriptaddress,showkeys,floatfix]{revtex4-1}
\usepackage{graphicx}
\usepackage{amssymb,amsfonts,amsmath}
\usepackage{psfrag}
\usepackage{natbib}

\begin{document}

\title{Fluctuating shells under pressure}

\author{Jayson Paulose}
\affiliation{Harvard School of Engineering and Applied Sciences, Cambridge MA 02138, USA}
\author{Gerard A. Vliegenthart}
\affiliation{Institute for Advanced Simulation, Forschungszentrum J\"ulich, D-52425 J\"ulich, Germany}
\author{Gerhard Gompper}
\affiliation{Institute for Advanced Simulation, Forschungszentrum J\"ulich, D-52425 J\"ulich, Germany}
\affiliation{Institute of Complex Systems, Forschungszentrum J\"ulich, D-52425 J\"ulich, Germany}
\author{David R. Nelson}
\affiliation{Department of Physics, Harvard University, Cambridge MA 02138, USA}

\begin{abstract} 
Thermal fluctuations strongly modify the large length-scale elastic behavior of crosslinked membranes, giving rise to scale-dependent elastic moduli. While thermal effects in flat membranes are well understood, many natural and artificial microstructures are modeled as thin elastic {\it shells}. Shells are distinguished from flat membranes by their nonzero curvature, which provides a size-dependent coupling between the in-plane stretching modes and the out-of-plane undulations. In addition, a shell can support a pressure difference between its interior and exterior. Little is known about the effect of thermal fluctuations on the elastic properties of shells. Here, we study the statistical mechanics of shape fluctuations in a pressurized spherical shell using perturbation theory and Monte Carlo computer simulations, explicitly including the effects of curvature and an inward pressure. We predict novel properties of fluctuating thin shells under point indentations and pressure-induced deformations. The contribution due to thermal fluctuations increases with increasing ratio of shell radius to thickness, and dominates the response when the product of this ratio and the thermal energy becomes large compared to the bending rigidity of the shell. Thermal effects are enhanced when a large uniform inward pressure acts on the shell, and diverge as this pressure approaches the classical buckling transition of the shell. Our results are relevant for the elasticity and osmotic collapse of microcapsules.

\end{abstract}


\maketitle

The elastic theory of thin plates and shells~\cite{landau_elasticity_1986}, a subject over a century old, has recently found new applications in understanding the mechanical properties of a wide range of natural and artificial structures at microscopic length scales. The mechanical properties of viral capsids~\cite{ivanovska_bacteriophage_2004,michel_nanoindentation_2006,klug_failure_2006}, red blood cells~\cite{park_measurement_2010}, and hollow polymer and polyelectrolyte capsules~\cite{gao_elasticity_2001, gordon_self-assembled_2004,lulevich_elasticity_2004,elsner_mechanical_2006,zoldesi_elastic_2008} have been measured and interpreted in terms of elastic constants of the materials making up these thin-walled structures. Theoretically, models that quantify the deformation energy of a two-dimensional membrane have been used to investigate the shapes of viral capsids~\cite{lidmar_virus_2003,nguyen_elasticity_2005,nguyen_continuum_2006} and their expected response to point forces and pressures~\cite{vliegenthart_mechanical_2006, buenemann_mechanical_2007,buenemann_elastic_2008,siber_stability_2009}, as well as shape transitions of pollen grains~\cite{katifori_foldable_2010}. 

Like its counterparts in other areas of science, such as fluid dynamics and the theory of electrical conduction in metals, thin shell theory aims to describe the physics of slowly varying disturbances in terms of a few macroscopic parameters, such as the shear viscosity of incompressible fluids and the electrical conductivity of metals.    Despite such venerable underpinnings as the Navier-Stokes equations and Ohm's law, these hydrodynamic theories can break down, sometimes in spectacular ways. For example, it is know from mode coupling theory~\cite{pomeau_time_1975} and from renormalization group calculations~\cite{forster_large-distance_1977} that thermal fluctuations cause the shear viscosity of incompressible fluids to diverge logarithmically with system size in a two-dimensional incompressible fluid. In the theory of electrical conduction, quenched disorder due to impurities coupled with interactions between electrons lead to a dramatic breakdown of Ohm's law in thin films and one-dimensional wires at low temperatures, with a conductance that depends on the sample dimensions~\cite{lee_disordered_1985}.

\begin{figure}
    \centering
    \includegraphics[width=87mm]{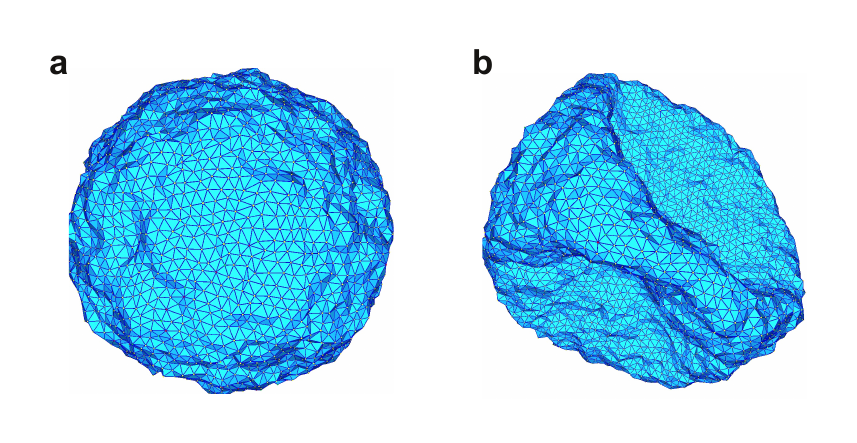}
    \caption{\textbf{Simulated thermally fluctuating shells.} \textbf{(a)} Triangulated shell with 5530 points separated by average nearest-neighbor distance $r_{0}$  with Young's modulus $Y=577\epsilon/r_{0}^{2}$ and bending rigidity $\kappa=50\epsilon$ at temperature $k_\text{B}T=20\epsilon$, where $\epsilon$ is the energy scale of the Lennard-Jones potential used to generate the disordered mesh. \textbf{(b)} Same as in (a) with external pressure $p=0.5p_{\mathrm{c}}$, where $p_{\mathrm{c}}$ is the classical buckling pressure. The thermally excited shell has already buckled under pressure to a shape with a much smaller enclosed volume than in (a).} \label{fig_schematic}
\end{figure}

Even more dramatic breakdowns of linear response theory can arise in thin plates and shells. Unlike the macroscopic shell structures of interest to civil engineers, thermal fluctuations can strongly influence structures with size of order microns, since  the elastic deformation energies of extremely thin membranes (with nanoscale thicknesses) can be of the order of the thermal energy $k_\text{B}T$ (where $k_\text{B}$ is the Boltzmann constant and $T$ the temperature) for typical deformations. The statistical mechanics of \emph{flat} solid plates and membranes (\emph{i.e.} membranes with no curvature in the unstrained state) has been studied previously (see~\cite{nelson_statistical_1988, bowick_statistical_2001} and references therein). Thermal fluctuations lead to \emph{scale-dependent} elastic moduli for flat membranes, causing the in-plane elastic moduli to vanish at large length scales while the bending rigidity diverges~\cite{nelson_fluctuations_1987, aronovitz_fluctuations_1989}. These anomalies arise from the the nonlinear couplings between out-of-plane deformations (transverse to the plane of the undeformed membrane) and the resultant in-plane strains, which are second order in the out-of-plane displacements.

Much less is known about spherical shells subject to thermal fluctuations (Fig.~1a). In fact, the coupling between in-plane and out-of-plane modes is significantly different. Geometry dictates that a closed spherical shell cannot be deformed without stretching; as a result, out-of-plane deformations provide a \emph{first} order contribution to the in-plane strain tensor~\cite{landau_elasticity_1986}. This introduces new nonlinear couplings between in-plane and out-of-plane deformations, which are forbidden by symmetry in flat membranes. We can also consider the buckling of spherical shells under uniform external pressure, which has no simple analogue for plates (Fig.~1b). An early exploration with computer simulations combined an analysis of the elastic energy due to the linear strain contributions of a spherical membrane with the nonlinear corrections from flat membranes to suggest new scaling behavior for thermally fluctuating spherical membranes~\cite{zhang_scaling_1993}. However, an important nonlinear coupling triggered by the curved background metric was not considered, nor was the effect of an external pressure investigated. Here, we study the mechanics of fluctuating spherical shells using perturbation theory and numerical simulations, taking into account the nonlinear couplings introduced by curvature as well as the effects of a uniform external pressure.

\section*{Results and discussion}
\subsection*{Elastic energy of a thin shell}
The elastic energy of a deformed spherical shell of radius $R$ is calculated using \emph{shallow-shell theory}~\cite{koiter_stability}. This approach considers a shallow section of the shell, small enough so that slopes measured relative to the section base are small.
The in-plane displacements of the shallow section are parametrized by a two-component phonon field $u_i(\mathbf{x})$, $ i={1,2}$; the out-of-plane displacements are described by a field $f(\mathbf{x})$ in a coordinate system $\mathbf{x}=(x_1,x_2)$ tangent to the shell at the origin. We focus on \emph{amorphous} shells, with uniform elastic properties, and can thus neglect the effect of the 12 inevitable disclinations associated with crystalline order on the surface of a sphere~\cite{lidmar_virus_2003}. In the presence of an external pressure $p$ acting inward, the elastic energy for small displacements in terms of the bending rigidity $\kappa$ and Lam\'e coefficients $\mu$ and $\lambda$ reads (see {\it Supplementary Information} for details):
\begin{equation}
G=\int d^2x\,\left[\frac{\kappa}{2} (\nabla^2 f)^2 +\mu u_{ij}^2 + \frac{\lambda}{2} u_{kk}^2-pf\right],
\end{equation} 
where the nonlinear strain tensor is 
\begin{equation}
u_{ij}(\mathbf{x})=\frac{1}{2}\left(\partial_i u_j+\partial_j u_i +\partial_i f \partial_j f\right)-\delta_{ij}\frac{f}{R}.
\end{equation}
Here, $d^{2}x \equiv \sqrt{g} dx_{1}dx_{2}$, where $g$ is the determinant of the metric tensor associated with the spherical background metric. Within shallow shell theory, $g \approx 1$ (see {\it Supplementary Information}).

If we represent the normal displacements in the form $f(\mathbf{x}) = f_0 + f^\prime(\mathbf{x})$, where $f_0$ represents the uniform contraction of the sphere in response to the external pressure, and $f^\prime$ is the deformation with reference to this contracted state so that $\int d^2x f^\prime =0$, then the energy is quadratic in fields $u_1$, $u_2$ and $f_0$. These variables can be eliminated in a functional integral of $\exp(-G[f^\prime, f_0,u_1,u_2]/k_\mathrm{B}T)$  by Gaussian integration (see {\it Supplementary Information} for details). The effective free energy $G_{\mathrm{eff}}$ which results is the sum of a harmonic part $G_0$ and an anharmonic part $G_1$ in the remaining variable $f^\prime(\mathbf{x})$:
\begin{eqnarray}
\label{eqn_effectivef_pressure}
G_0 & = & \frac{1}{2}\int d^2 x\left[\kappa(\nabla^2 f^\prime)^2-\,\frac{pR}{2} |\nabla f^\prime|^2+\frac{Y}{R^2} {f^\prime}^2\right],\\
G_1& = &\frac{Y}{2}\int d^2 x\left[\left(\frac{1}{2}P^\mathrm{T}_{ij}\partial_i f^\prime \partial_j f^\prime\right)^2-\frac{f^\prime}{R}P^\mathrm{T}_{ij}\partial_i f^\prime \partial_j f^\prime \right]. \nonumber
\end{eqnarray}
where $Y = 4\mu(\mu+\lambda)/(2\mu+\lambda)$ is the two-dimensional Young modulus and $P^\mathrm{T}_{ij}=\delta_{ij}-\partial_i \partial_j/\nabla^2$ is the transverse projection operator.  The ``mass'' term $Y({f^\prime}/R)^2$ in the harmonic energy functional reflects the coupling between out-of-plane deformation and in-plane stretching due to curvature, absent in the harmonic theory of flat membranes (plates). The cubic interaction term with a coupling constant $-Y/2R$ is also unique to curved membranes and is prohibited by symmetry for flat membranes. These terms are unusual because they have system-size-dependent coupling constants. Note that an inward pressure ($p >0$) acts like a negative $R$-dependent surface tension in the harmonic term. As required, the effective elastic energy of fluctuating flat membranes is retrieved for $R\to \infty$ and $p = 0$. In the following, we exclusively use the field $f^\prime(\mathbf{x})$ and thus drop the prime without ambiguity.

When only the harmonic contributions are considered, the equipartition result for the thermally generated Fourier components $f_\mathbf{q} = \int d^2x \,f(\mathbf{x})\exp(i\mathbf{q}\cdot\mathbf{x})$ with two-dimensional wavevector $\mathbf{q}$ are
\begin{equation} \label{eqn_corrfn_gaussian}
\langle f_\mathbf{q}f_\mathbf{q^\prime} \rangle_0 = \frac{Ak_\mathrm{B}T \delta_{\mathbf{q},\mathbf{-q^\prime}}}{\kappa q^4 -\frac{pR}{2}q^2+ \frac{Y}{R^2}}.
\end{equation}
where $A$ is the area of integration in the $(x_1,x_2)$ plane. Long-wavelength modes are restricted by the finite size of the sphere, \emph{i.e.} $q \gtrsim 1/R$. In contrast to flat membranes for which the amplitude of long-wavelength ($q \to 0$) modes diverges as $k_\text{B}T/(\kappa q^4)$, the coupling between in-plane and out-of-plane deformations of curved membranes cuts off fluctuations with wavevectors smaller than a characteristic inverse length scale~\cite{zhang_scaling_1993}: $$q^* = (\ell^*)^{-1} = \left(\frac{Y}{\kappa R^2}\right)^{1/4} \equiv  \frac{\gamma^{1/4}}{R},$$ where we have introduced the dimensionless
{\it F\"oppl-von K\'arm\'an number}
$\gamma = YR^2/\kappa$ \cite{lidmar_virus_2003}. We focus here on the case $\gamma \gg 1$, so $\ell^{*} \ll R$. As  $p$ approaches $p_\mathrm{c} \equiv 4\sqrt{\kappa Y}/R^2$, the modes with $q = q^*$ become unstable and their amplitude diverges. This corresponds to the well-known buckling transition of spherical shells under external pressure~\cite{koiter_stability}. When $p > p_\mathrm{c}$, the shape of the deformed shell is no longer described by small deformations from a sphere, and the shallow shell approximation breaks down.

\begin{figure*}
    \centering{}
    \includegraphics[width=170mm]{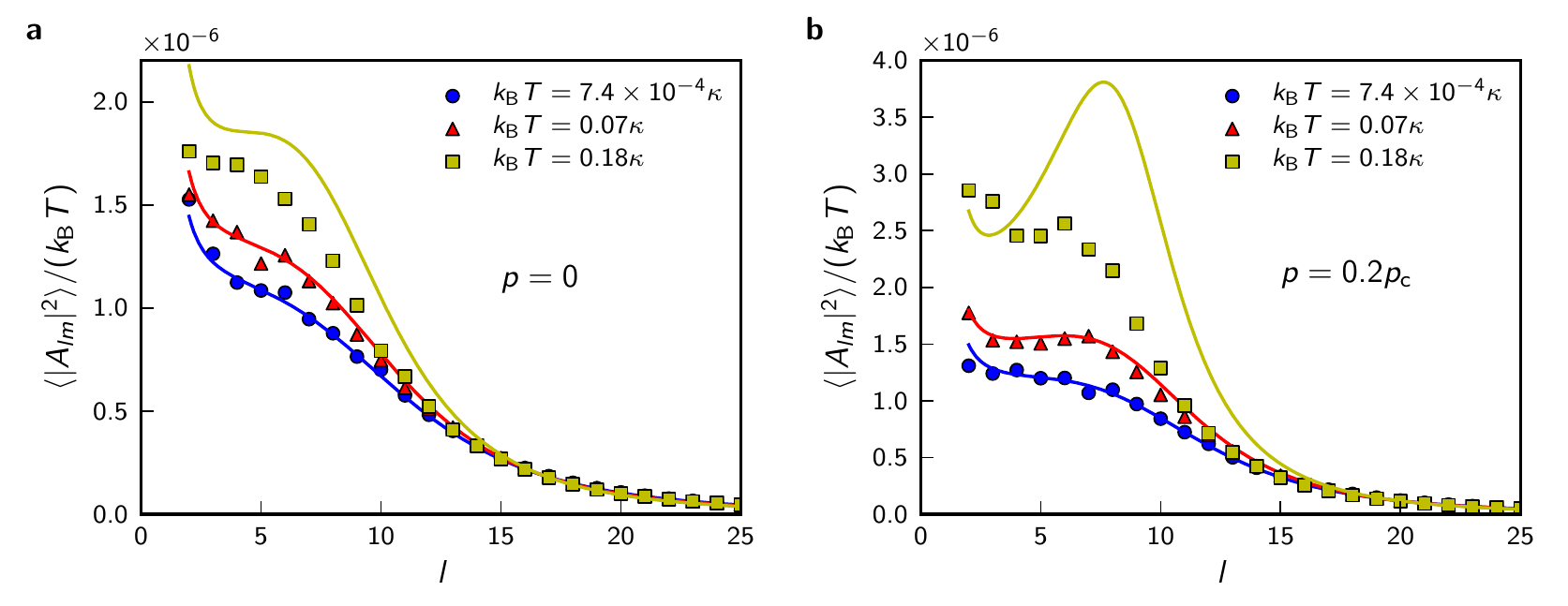}
    \caption{\textbf{Fluctuation spectrum in spherical harmonics.} Spherical harmonic amplitude of the shape fluctuations of elastic shells plotted against the dimensionless spherical wavenumber $l$ for a shell with $R=40r_{0}, Y=577\epsilon/r_{0}^{2}$ and $\kappa=50\epsilon$ at temperatures $k_\text{B}T/\kappa = 7.4\times 10^{-4}$ (blue), 0.07 (red) and 0.18 (yellow). The fluctuation amplitudes are scaled by $k_{\text{B}}T$ so that the spectra at different temperatures would coincide in the harmonic approximation. Each subfigure corresponds to a different value of the external pressure: $p=0$ (\textbf{a}) and $p=0.2p_{\text{c}}$ (\textbf{b}). The symbols are from Monte Carlo simulations, and the solid lines are the theoretical prediction, Eq.~\ref{eqn-almsurf}, using the renormalized elastic constants from perturbation theory (Eqs.~\ref{eqn_smallpyr}--\ref{eqn_smallpkr}), except for the lowest temperature, where the bare elastic constants are used since the anharmonic effects are negligible.}\label{fig_sphfl}
\end{figure*}

\subsection*{Anharmonic corrections to elastic moduli}

The anharmonic part of the elastic energy, neglected in the analysis described above, modifies the fluctuation spectrum by coupling Fourier modes at different wavevectors. Upon rescaling all lengths by $\ell^*$, it can be shown that the size of 
anharmonic contributions to $\langle |f_{\mathbf{q}}|^2 \rangle$ is set by
the dimensionless quantities $k_\text{B}T \sqrt{\gamma}/\kappa$ and $p/p_\text{c}$. The correlation function including the anharmonic terms in Eq.~\ref{eqn_effectivef_pressure}  is given by the Dyson equation,
\begin{equation} \label{eqn_dyson}
\langle |f_\mathbf{q}|^2 \rangle = \frac{1}{\langle |f_\mathbf{q}|^2 \rangle_0^{-1}-\Sigma(\mathbf{q})}
\end{equation}
where $\Sigma(\mathbf{q})$ is the self-energy, which we evaluate to one-loop order using perturbation theory. While $\langle |f_\mathbf{q}|^2 \rangle$ can be numerically evaluated at any $\mathbf{q}$, an approximate but concise description of the fluctuation spectrum is obtained by expanding the self-energy up to order $q^4$ and defining renormalized values $Y_{\scriptscriptstyle\mathrm{R}}$, $\kappa_{\scriptscriptstyle\mathrm{R}}$ and $p_{\scriptscriptstyle\mathrm{R}}$ of the Young's modulus, bending rigidity and pressure, from the coefficients of the expansion:
\begin{equation}
Ak_\mathrm{B}T\langle|f_\mathbf{q\rightarrow 0}|^2\rangle^{-1} \equiv \kappa_{\scriptscriptstyle\mathrm{R}} q^4 - \frac{p_{\scriptscriptstyle\mathrm{R}}R}{2} q^2+\frac{Y_{\scriptscriptstyle\mathrm{R}}}{R^2} + O(q^6).
\end{equation}
To lowest order in $k_\mathrm{B}T/\kappa$ and $p/p_\text{c}$ we obtain the approximate expressions (see {\it Supplementary Information} for details)
\begin{equation} \label{eqn_smallpyr}
Y_{\scriptscriptstyle\mathrm{R}} \approx  Y \left[1  -\frac{3}{256}\frac{k_\mathrm{B}T}{\kappa}\sqrt{\gamma}\left(1+\frac{4}{\pi}\frac{p}{p_{\text{c}}}\right)\right],
\end{equation}
\begin{equation} \label{eqn_genpressure}
p_{\scriptscriptstyle\mathrm{R}} \approx p +\frac{1}{24\pi}\frac{k_\mathrm{B}T}{\kappa}p_\text{c}\sqrt{\gamma}\left(1+\frac{63\pi}{128}\frac{p}{p_{\text{c}}}\right) ,
\end{equation}
and
\begin{equation}\label{eqn_smallpkr}
\kappa_{\scriptscriptstyle\mathrm{R}} \approx \kappa\left[1 +\frac{61}{4096}\frac{k_\mathrm{B}T}{\kappa}\sqrt{\gamma}\left(1-\frac{1568}{915\pi}\frac{p}{p_{\text{c}}}\right)\right].
\end{equation}
(See {\it Supplementary Information} for details of the calculation and the complete dependence on $p/p_\text{c}$.) Thus the long-wavelength deformations of a thermally fluctuating shell are governed by a smaller effective Young's modulus, a larger effective bending rigidity, and a nonzero negative surface tension even when the external pressure is zero.  At larger $p/p_\text{c}$, however, both the Young's modulus and the bending modulus fall compared to their zero temperature values, and the negative effective surface tension determined by $p_{\scriptscriptstyle\mathrm{R}}$ gets very large. The complete expressions for the effective elastic parameters, including the full $p/p_\text{c}$-dependence, show that all corrections diverge as $p/p_\text{c} \to 1$.    Furthermore, the effective elastic constants are not only temperature-dependent, but also system size-dependent, since $\sqrt\gamma\propto R$. Although the corrections are formally small for $k_{\text{B}}T \ll \kappa$, they nevertheless diverge as $R \to \infty$! The thermally generated surface tension, strong dependence on external pressure, and size dependence of elastic constants are unique to spherical membranes, with no analogue in planar membranes.

\subsection*{Simulations of thermally fluctuating shells}

\begin{figure*}
    \centering{}
        \includegraphics[width=170mm]{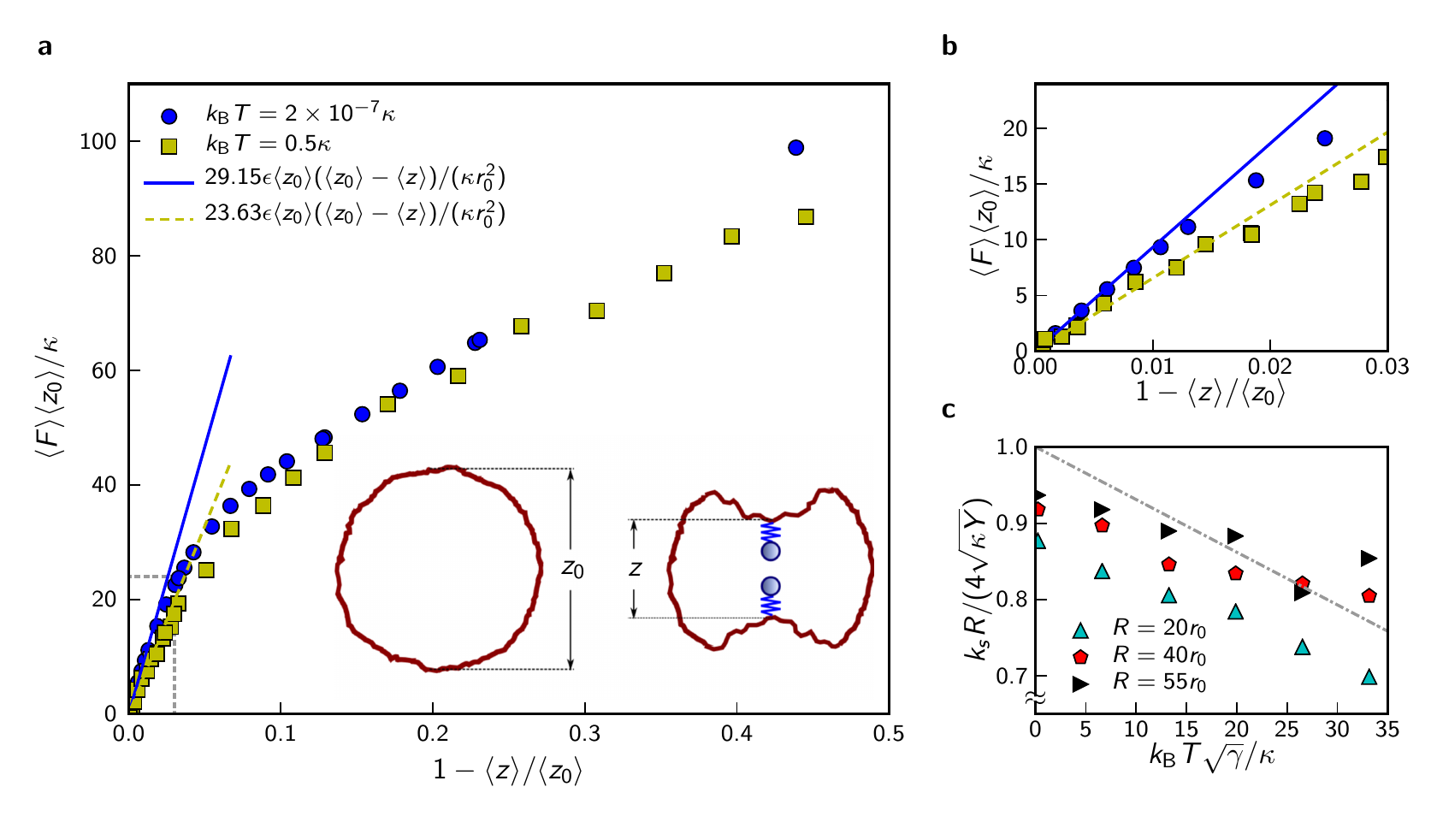}
    \caption{\textbf{Temperature dependence of response to point forces.} (\textbf{a}) Force-compression curves for simulations of indented shells (symbols) with $R=20 r_{0}$, $Y=577\epsilon/r_{0}^{2}$ and $\kappa = 50\epsilon$ at low ($k_\text{B}T/\kappa = 2\times 10^{-7}$) and high ($k_\text{B}T/\kappa = 0.5$) temperature. The lines show the expected linear response at small deformations with the spring constant $k_\text{s}$ measured independently from fluctuations in $z_{0}$ ($k_\text{s} = 29.15\epsilon/r_{0}^{2}$ for $k_\text{B}T/\kappa = 2\times 10^{-7}$, $k_\text{s} = 23.63\epsilon/r_{0}^{2}$ for $k_\text{B}T/\kappa = 0.5$). For indentation depths larger than $1-\langle z \rangle/\langle z_{0}\rangle \approx 0.05$, the regions around the poles become inverted and the response becomes nonlinear. Inset: schematic showing the definition of $z_{0}$ (the pole-to-pole distance in the absence of indentations) and $z$ (pole-to-pole distance following an indentation imposed by harmonic springs  whose free ends are brought close together) for a snapshot of the fluctuating shell. (\textbf{b}) Blow-up of the boxed region near the origin in (\textbf{a}), highlighting the linear response regime. (\textbf{c}) Spring constants extracted from fluctuations for shells with three different radii as a function of temperature, rescaled by the classical result for linear response of thin shells at zero temperature. The dashed line shows the perturbation theory prediction, Eq.~\ref{eqn0sprconst}. The low-temperature spring constant deviates from the classical result due to a finite mesh size effect which falls with increasing $R$ (increasing mesh size). } \label{fig_indent}
\end{figure*}

We complement our theoretical calculations with Monte Carlo simulations of randomly triangulated spherical shells with discretized bending and stretching energies that translate directly into a macroscopic 2D shear modulus $Y$ and a bending ridigity $\kappa$~\cite{vliegenthart_forced_2006,vliegenthart_compression_2011}. (Details are provided in {\it Materials and Methods}.) Here we study shells with $600 < \gamma < 35000$ and $2\times 10^{-6} <k_\text{B}T/\kappa < 0.5$. The anharmonic effects are negligible at the low end of this temperature range.

The fluctuation spectra of the simulated spherical shells are evaluated using an expansion of the radial displacement field in spherical harmonics~\cite{gompper_random_1996}. The radial position of a node $i$ at angles ($\phi,\theta$) can be  written as $r_i(\phi,\theta)=\widetilde{R_0}+f(\phi,\theta)$
with $\widetilde{R_0}$ the average radius of the fluctuating vesicle.
The function $f(\phi,\theta)$ can be expanded in (real) spherical harmonics
\begin{equation} \label{eqn0sphharm}
f(\phi,\theta)=R\sum_{l=0}^{l_M}\sum_{m=-l}^{m=l} A_{lm}Y_{lm}(\phi,\theta)
\end{equation}
where $l_M$ is the large wavenumber cutoff determined by the number of nodes in the lattice 
$(l_M+1)^2=N$ \cite{gompper_random_1996}. The theoretical prediction for the  fluctuation spectrum including anharmonic effects is ({\it Supplementary Information})
\begin{equation}
    \begin{split}
        k_\text{B}T\langle |A_{lm}|^{2}\rangle^{-1} \approx &\kappa_{\scriptscriptstyle\mathrm{R}}(l+2)^{2}(l-1)^{2} -p_{\scriptscriptstyle\mathrm{R}}R^{3}\left[1+\frac{l(l+1)}{2}\right] \\
        &+Y_{\scriptscriptstyle\mathrm{R}}R^{2}\left[\frac{3(l^{2}+l-2)}{3(l^{2}+l)-2}\right]. \label{eqn-almsurf}
    \end{split}
\end{equation}
Fig.~\ref{fig_sphfl} displays our theoretical and simulation results for the fluctuation spectrum. At the lowest temperature (corresponding to $k_\text{B}T\sqrt\gamma/\kappa \approx 0.1 \ll 1$), the spectrum is well-described by the bare elastic parameters $Y$, $\kappa$ and $p$. At the intermediate temperature ($k_\text{B}T\sqrt\gamma/\kappa \approx 10$) anharmonic corrections become significant, enhancing the fluctuation amplitude for some values of $l$ by about 20\%--40\% compared to the purely harmonic contribution. At this temperature, one-loop perturbation theory successfully describes the fluctuation spectrum. However, at the highest temperature simulated ($k_\text{B}T\sqrt\gamma/\kappa \approx 24$), the anharmonic corrections observed in simulations approach 50\% of the harmonic contribution at zero pressure and over 100\% for the pressurized shell. With such large corrections, we expect that higher-order terms in the perturbation expansion contribute significantly to the fluctuation spectrum and the one-loop result overestimates the fluctuation amplitudes.

Similarly, thermal fluctuations modify the mechanical response when a shell is deformed by a deliberate point-like indentation. In experiments, such a deformation is accomplished using an atomic force microscope~\cite{ivanovska_bacteriophage_2004,elsner_mechanical_2006}. In our simulations, two harmonic springs are attached to the north and south pole of the shell. By changing the position of the springs the depth of the indentation can be varied (Fig.~\ref{fig_indent}a, inset). The thermally averaged pole-to-pole distance $\langle z \rangle$ is measured and compared to its average value in the absence of a force, $\langle z_{0} \rangle$. For small deformations, the relationship between the force applied at each pole and the corresponding change in pole--pole distance is spring-like with a spring constant $k_{\text{s}}$: $\langle F \rangle \equiv k_{\text{s}} (\langle z_{0} \rangle - \langle z \rangle)$. The spring constant is related to the amplitude of thermal fluctuations in the normal displacement field in the \emph{absence} of forces by (see {\it Supplementary Information} for the detailed derivation)
\begin{equation} \label{eqn0ksfluct}
    k_{\text{s}} = \frac{k_\text{B}T}{2\langle [f(\mathbf{x})]^{2} \rangle} \approx \frac{k_\text{B}T}{\langle z_{0}^{2}\rangle - \langle z_{0}\rangle^{2}}.
\end{equation}
This fluctuation-response relation is used to measure the temperature dependence of $k_{\text{s}}$ from simulations on fluctuating shells with no indenters. At finite temperature, anharmonic effects computed above make this spring constant both size- and temperature-dependent:
\begin{equation} \label{eqn0sprconst}
    k_{\text{s}} \approx \frac{4\sqrt{\kappa Y}}{R}\left[1-0.0069\frac{k_\text{B}T}{\kappa} \sqrt\gamma\right].
\end{equation}

Fig.~\ref{fig_indent}a shows the force-compression relation for a shell with $R = 20 r_0$ and dimensionless temperatures $k_\text{B}T\sqrt\gamma/\kappa = 1.36 \times 10^{-4}$ and $k_\text{B}T\sqrt\gamma/\kappa = 34$. The linear response near the origin (Fig.~\ref{fig_indent}b) is very well described by $k_{\text{s}}$ measured indirectly from the fluctuations in $z_{0}$ at each temperature, Eq.~\ref{eqn0ksfluct}. The thermal fluctuations lead to an appreciable 20\% reduction of the spring constant for this case. Measuring spring constants over a range of temperatures (Fig.~\ref{fig_indent}c) confirms that the shell response softens as the temperature is increased, in agreement with the perturbation theory prediction. We note, however, a small but systematic shift due to the finite mesh size of the shells, an approximately 5\% effect for the largest systems simulated here. At the higher temperatures ($k_\text{B}T\sqrt\gamma/\kappa >20$), the measured spring constants deviate from the perturbation theory prediction, once again we believe due to the effect of higher-order terms.

\begin{figure}
    \centering{}
    \includegraphics[width=87mm]{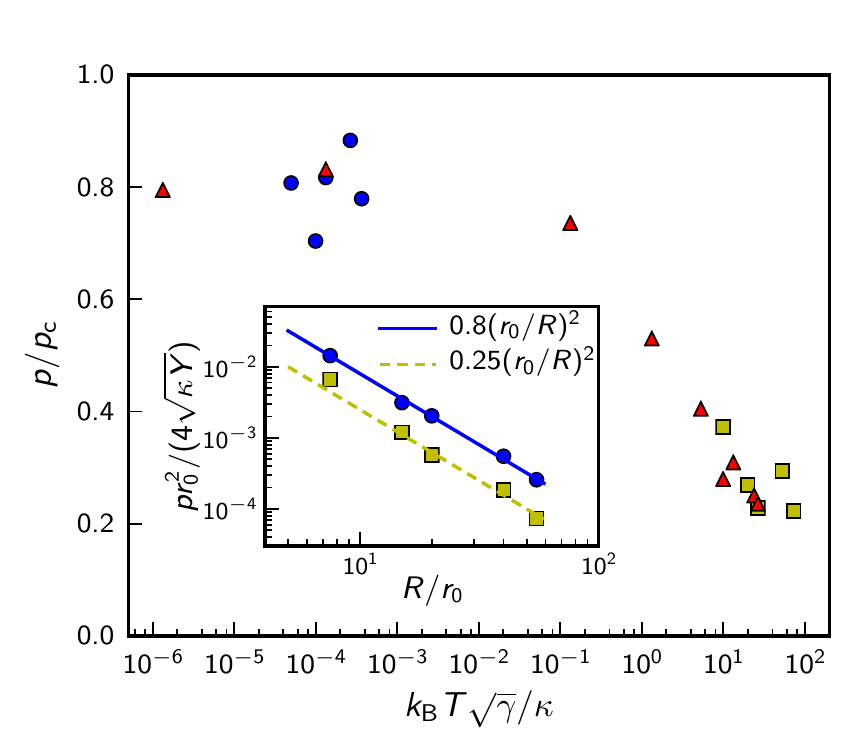}
    \caption{\textbf{Temperature dependence of the buckling pressure.} Buckling pressure for simulated shells at various radii and temperatures, normalized by the \emph{classical} (\emph{i.e.} zero temperature) critical buckling pressure $p_\text{c}$ for perfectly uniform, zero temperature shells with the same parameters. For all shells, $Yr_{0}^{2}/\kappa = 11.54$. In separate sets of symbols, we either vary the shell radius over the range $7.5 \leq R/r_{0} \leq 55$ while keeping the temperature constant ($k_{\text{B}}T = 2\times 10^{-6}\kappa$, blue circles; $k_{\text{B}}T = 0.4\kappa$, yellow squares) or vary the temperature over the range $2\times 10^{-8} \leq k_{\text{B}}T/\kappa \leq 0.4$ while keeping the radius constant at $R=20 r_{0}$ (red triangles). The parameter $k_{\text{B}}T\sqrt\gamma/\kappa$ sets the strength of anharmonic corrections for thermally fluctuating shells. The inset shows the $1/R^{2}$ dependence of the buckling pressure as the radius is varied, for shells at low and high temperature.} \label{fig_critpres}
\end{figure}

We also simulate the buckling of thermally excited shells under external pressure. When the external pressure increases beyond a certain value (which we identify as the renormalized buckling pressure), the shell collapses from a primarily spherical shape (Fig.~1a) to a shape with one or more large volume-reducing inversions (Fig.~1b). For zero temperature shells, this buckling is associated with the appearance of an unstable deformation mode in the fluctuation spectrum. At finite temperature, the appearance of a mode with energy of order $k_\text{B}T$ is sufficient to drive buckling. Anharmonic contributions, strongly enhanced by an external pressure, also reduce the effective energy associated with modes in the vicinity of $q^{*}$ primarily due to the enhanced negative effective surface tension $p_{\scriptscriptstyle\mathrm{R}}R/2$ (see Eq.~\ref{eqn_genpressure}). As a result, unstable modes arise at lower pressures and we expect thermally fluctuating shells to collapse at pressures below the classical buckling pressure $p_\text{c}$. This is confirmed by simulations of pressurized shells (Fig.~\ref{fig_critpres}). When anharmonic contributions are negligible ($k_\text{B}T\sqrt\gamma/\kappa \ll 1$), the buckling pressure observed in simulations is only  $\sim 80\%$ of the theoretical value because the buckling transition is highly sensitive to the disorder introduced by the random mesh. Relative to this low temperature value, the buckling pressure is reduced significantly when $k_\text{B}T\sqrt\gamma/\kappa$ becomes large.

\subsection*{Conclusion and outlook}
In summary, we have demonstrated that thermal corrections to the elastic response become significant when $k_\text{B}T\sqrt{\gamma}/\kappa \gg 1$ and that first-order corrections in $k_\text{B}T/\kappa$ already become inaccurate when $k_\text{B}T\sqrt{\gamma}/\kappa \gtrsim 20$. Human red blood cell (RBC) membranes are known examples of curved solid structures that are soft enough to exhibit thermal fluctuations. Typical measured values of the shear and bulk moduli of RBC membranes correspond to $Y\approx 25$ $\mu$N/m~\cite{park_measurement_2010,waugh_thermoelasticity_1979}, while reported values of the bending rigidity $\kappa$ vary widely from 6 $k_\text{B}T$ to 40 $k_\text{B}T$~\cite{park_measurement_2010,evans_bending_1983}. Using an effective radius of curvature $R \approx 7$ $\mu$m~\cite{park_measurement_2010} gives $k_\text{B}T\sqrt{\gamma}/\kappa$ in the range 2--35. Thus, RBCs could be good candidates to observe our predicted thermal effects, provided their bending rigidity is in the lower range of the reported values.

For continuum shells fabricated from an elastic material with a 3D Young's modulus $E$, thickness $h$ and typical Poisson ratio $\approx 0.3$, $k_\text{B}T\sqrt\gamma/\kappa \approx 100Rk_\text{B}T/(Eh^4)$. Hence very thin shells with a sufficiently high radius-to-thickness ratio ($R/h$) \emph{must} display significant thermal effects. Polyelectrolyte~\cite{elsner_mechanical_2006} and protein-based~\cite{hermanson_engineered_2007} shells with $R/h \approx 10^3$ have been fabricated, but typical solid shells have a bending rigidity $\kappa$ several orders of magnitude higher than $k_\text{B}T$ unless $h \lesssim 5$ nm. Microcapsules of 6 nm thickness fabricated from reconstituted spider silk~\cite{hermanson_engineered_2007} with $R \approx 30$ $\mu\mathrm{m}$ and $E \approx 1$ GPa  have $k_\text{B}T\sqrt{\gamma}/\kappa \approx 3$, and could exhibit measurable anharmonic effects.

Thermal effects are particularly pronounced under finite external pressure---an indentation experiment carried out at $p = p_\mathrm{c}/2$ on the aforementioned spider silk capsules would show corrections of 10\% from the classical zero-temperature theory. For similar capsules with half the thickness, perturbative corrections at $p=p_\mathrm{c}/2$ are larger than 100\%, reflecting a drastic breakdown of shell theory because of thermal fluctuations. The breakdown of classical shell theory explored here points to the need for a renormalization analysis, similar to that carried out already for flat plates~\cite{statistical_1988}.

\appendix*

\section*{Materials and Methods}
\subsection*{Monte Carlo Simulations of randomly triangulated shells} \label{sec0sisimulation}
A random triangulation of radius $R_0$ is constructed by  distributing $N$ nodes on the surface of a sphere with the required radius. The first two of these nodes are fixed at the north and the south pole of the sphere whereas the positions of  the remaining $N-2$ nodes are randomized and equilibrated in a Monte Carlo simulation. During  this equilibration process the nodes interact via a steeply repulsive potential (the repulsive part of a Lennard Jones potential). After equilibration, when the energy has reached a constant value on average, the simulation is stopped and  the final configuration is `frozen'. The neighbours of all nodes are determined using a Delaunay triangulation \cite{renka_algorithm_1997}. The spherical configurations as well as the connection lists are used in further simulations.

In subsequent simulations nearest neighbours are permanently linked by a harmonic  potential giving rise to a total stretching energy \cite{seung_defects_1988}, 
\begin{equation}
    E_\text{s}=\frac{k}{2}\sum_{i,j} (|r_{ij}-r_{ij}^0|^2),
\end{equation}
where the sum runs over all pairs of nearest neighbours, $r_{ij}$ is the distance between two neighbours and $r_{ij}^0$ the equilibrium length of a spring.  The equilibrium length $r_{ij}^0$ is determined at the start of the simulation, when the shell is still perfectly spherical and thus the stretching energy  vanishes for the spherical shape. The spring constant $k$ is related to the two-dimensional Lam\'e coefficients  $\lambda=\mu=\sqrt{3}k/4$ and the two-dimensional Young modulus $Y=2 k/\sqrt{3}$ \cite{seung_defects_1988}.

The mean curvature (more precisely, twice the mean curvature) at node $i$ is discretized using \cite{gompper_random_1996,itzykson_discrete_1986,kohyama_budding_2003}
\begin{equation}
H_i = \frac{1}{\sigma_i}{\bf n}_i \cdot \sum_{j(i)}\frac{\sigma_{ij}}{l_{ij}}({\bf r}_i-{\bf r}_j)
\end{equation}
where ${\bf n}_i$ is the surface (unit) normal at node $i$ (the average normal of the faces surrounding node $i$), $\sigma_i=\sum_{j(i)}\sigma_{ij}l_{ij}$ is the area of the dual cell of node $i$, $\sigma_{ij}=l_{ij}[\cot{\theta_1} + \cot{\theta_2}]/2$ is the length of a bond in the dual lattice and $l_{ij}=|{\bf r}_i-{\bf r}_j|$ is the distance between the nodes $i$ and $j$.
The total curvature energy is,
\begin{equation}
E_\text{b} = \frac{\kappa}{2} \sum_i \sigma_i (H_i-H_0)^2
\end{equation}
with $\kappa$ the bending rigidity and $H_0$ the spontaneous curvature at node $i$. In all simulations $H_0={2/R_0}$ (since $H_i$ is twice the mean curvature). In the cases of elastic shells under pressure a term $P V$ is added to the Hamiltonian where $P$ is the external pressure and $V$ the volume of the shell.

Similar elastic networks with stretching and bending potentials have been studied in relation to the stability of membranes, icosahedral and spherical shells that contain defects \cite{lidmar_virus_2003,siber_stability_2009,vliegenthart_compression_2011,seung_defects_1988,gompper_triangulated-surface_2004,widom_soft_2007} or defect scars \cite{kohyama_budding_2003,bowick_interacting_2000,bausch_grain_2003,kohyama_defect_2007} as well as for the deformation of icosahedral viruses \cite{vliegenthart_mechanical_2006,buenemann_mechanical_2007,buenemann_elastic_2008} and the crumpling of elastic sheets~\cite{vliegenthart_forced_2006}.

Simulations are performed for shells of 
5530 ($R_0=20 \ r_0$), 22117 ($R_0=40 \ r_0$) and 41816 ($R = 55 \ r_{0}$) nodes. 
The Hookean spring constant and the bending rigidity are taken such that the shells have F\"oppl-von-K\'arm\'an numbers in the range $650<\gamma<35000$ and that the dimensionless temperature is in the range $2 \times 10^{-6}< k_\text{B}T/\kappa<0.5$. Monte Carlo production runs consist typically of $1.25 \times 10^8$ Monte Carlo steps where in a single Monte Carlo step an attempt is made to update the positions of all nodes once on average. Configurations were stored for analysis typically every $N_\text{samp}=2000$ Monte Carlo steps. For the largest system (41816 nodes), such a run took about 700 days of net CPU time spread over several simultaneous runs in a Linux cluster of Intel XEON X5355 CPUs. For the smaller shells, the computational time scaled down roughly linearly with system size.

\subsection*{The fluctuation spectrum from computer simulations}
For a particular configuration of a simulated shell, the coefficients $A_{lm}$ of the expansion of the radial displacements in spherical harmonics (Eq.~\ref{eqn0sphharm}) are determined by a least squares fit of the node positions to 
a finite number $l_M$ of 
(real) spherical harmonics. In practice we have used $l_M=26$ as the upper wavenumber cutoff for all simulations.  At each temperature and pressure, this procedure is repeated for about 10000 independent configurations and the results averaged to obtain the curves presented in Fig.~\ref{fig_sphfl}.

\subsection*{Simulations of shells indented by point-like forces}
To perform indentation simulations, two harmonic springs are attached to the north and south pole of the shell. This leads to an additional term in the Hamiltonian $V_\text{s}=k_\text{i}\left(z_\text{i}^\text{N} - z^\text{N} \right)^2/2+k_\text{i}\left(z_\text{i}^\text{S} - z^\text{S} \right)^2/2$ where $k_\text{i}=\kappa/r_{0}^{2}$ is the spring constant of the indenter.
Here, one end of the springs, at positions $z^\text{N}$ and $z^\text{S}$, is attached to the vertices at the north and south pole, respectively. The positions of the other end of the springs, at $z_\text{i}^\text{N}$ and $z_\text{i}^\text{S}$, are fixed externally and determine the indentation force and depth, as indicated in Fig.~S4.

By changing $z_\text{i}^\text{N}$ and $z_\text{i}^\text{S}$, the depth of the indentation can be varied. After the springs are fixed a certain distance apart, the thermally average pole-to-pole distance $\langle z \rangle$ is measured and compared to its value in the absence of a force, $\langle z_{0} \rangle$. The instantaneous force at the poles is calculated from the instantaneous extension of the harmonic springs after each $N_\text{samp}$ Monte Carlo steps; thermal averaging then determines the average corresponding to $\langle z \rangle$. This provides the force-indentation curves in Fig. 3(a--b). 

It is very difficult to unambiguously identify the linear regime in the force-indentation curves. Extracting the effective spring constant  of shell deformation $k_\text{s}$ from a linear fit in the small indentation region is subject to inaccuracies and sensitivity to the number of points included in fitting.  Instead, we extract the spring constants of thermally fluctuating shells by using a relation between $k_\text{s}$ and the fluctuations in $z_0$ (see {\it Supplementary Information} for derivation):
\begin{equation}
    k_{\text{s}} \approx \frac{k_\text{B}T}{\langle z_{0}^{2}\rangle - \langle z_{0}\rangle^{2}}.
\end{equation}
This procedure was used to measure the temperature-dependent spring constants in Fig. 3c.

\begin{acknowledgments}
It is a pleasure to acknowledge J. Hutchinson, F. Spaepen, Z. Zeravcic and A. Kosmrlj for helpful discussions. Work by JP and DRN was supported by the National Science Foundation via Grant DMR1005289 and through the Harvard Materials Research Science and Engineering Center through Grant DMR0820484.
\end{acknowledgments}

\clearpage

\onecolumngrid
\renewcommand{\theequation}{S\arabic{equation}}
\renewcommand{\thefigure}{S\arabic{figure}}
\renewcommand{\thesection}{S\arabic{section}}
\renewcommand{\thetable}{S\arabic{table}}
\setcounter{figure}{0}
\setcounter{equation}{0}
\setcounter{section}{0}
\setcounter{table}{0}

\begin{center}
{\LARGE Supplementary Information}
\end{center}

\section*{Fields and strains in shallow shell theory}

\begin{figure*}[h]
\centering
\includegraphics[width=3.42in]{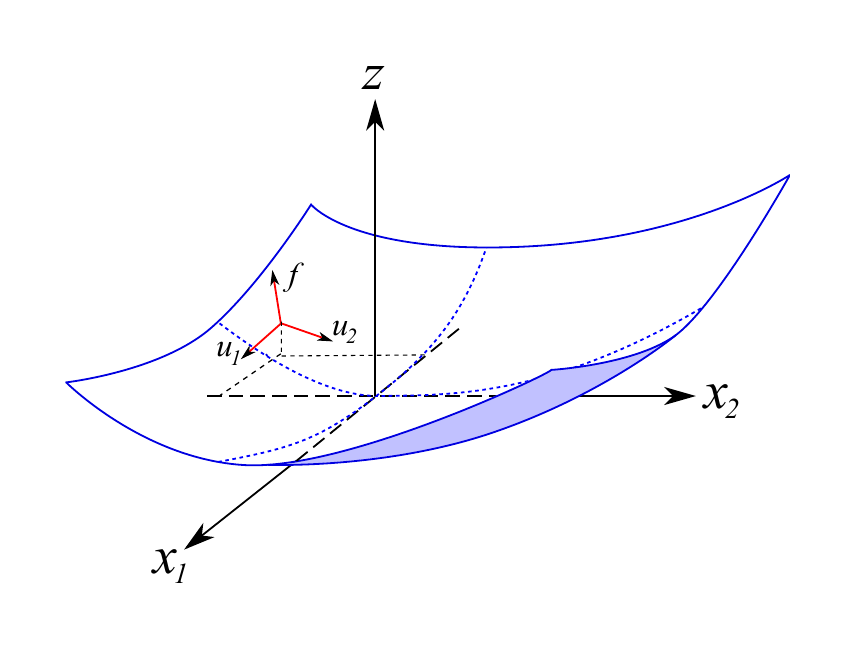}
\caption{The coordinate system in shallow shell theory. A section of the undeformed sphere is shown with the $(x_1,x_2)$ plane tangential to it at the origin. The red arrows show the directions into which displacements $u_1(x_1,x_2)$, $u_2(x_1,x_2)$ and $f(x_1,x_2)$ are decomposed at a particular point in the coordinate plane.} \label{si_fig_shallow}
\end{figure*}

We describe the deformations of the sphere using \emph{shallow shell theory} which we summarize here. We follow the presentation by Koiter and van der Heijden~\cite{si_koiter_stability}. A shallow section of the sphere is isolated and Cartesian coordinates $(x_1,x_2)$ are set up to define a plane that just touches the undeformed sphere at the origin and lies tangent to it; the $z$ axis is thus normal to the sphere at the origin (Fig.~S1). We use the Monge representation to parametrize the undeformed shell by its height $z=Z(x_1,x_2)$ above the plane, where $Z(x_1,x_2)$ is the undeformed state corresponding to a sphere of radius $R$ with its center located on the $z$-axis above the $(x_1,x_2)$ plane;
\begin{equation}
Z(x_1,x_2) = R\left(1-\sqrt{1-\frac{x_1^2}{R^2}-\frac{x_2^2}{R^2}}\right)
\end{equation}
The assumption in shallow shell theory is that the section of the shell under consideration is small enough that slopes $\partial_1 Z \sim x_1/R$ and $\partial_2 Z \sim x_2/R$ measured relative to the $(x_1,x_2)$ plane are small. (Partial derivatives are denoted by $\partial/\partial x_i \equiv \partial_i$.) Then the undeformed state is approximately parabolic in $x_1$ and $x_2$, 
\begin{equation}
Z(x_1,x_2) \approx \frac{x_1^2+x_2^2}{2R}.
\end{equation}

Deformations from this initial state are quantified \emph{via} a local normal displacement $f(x_1,x_2)$  perpendicular to the undeformed surface and tangential displacements $u_1(x_1,x_2)$ and $u_2(x_1,x_2)$ within the shell along the projections of the $x_1$ and $x_2$ axes on the sphere respectively. In terms of these fields, a point $(x_1, x_2, Z(x_1,x_2))$ in the undeformed state moves to $\left(x_1+u_1- f \partial_1 Z, x_2+u_2- f \partial_2 Z, Z+f\right)$ to lowest order in the slopes $\partial_i Z=x_i/R$.  The strain tensor is defined by the relation between the length $ds^\prime$ of a line element in the deformed state and the corresponding line element length $ds$ in the undeformed state~\cite{si_landau_elasticity_1986}: 
\begin{equation}
(ds^\prime)^2 = ds^2 + 2u_{ij}dx_idx_j.
\end{equation}
With this definition and neglecting terms of order $(\partial_i Z)^2$ and their derivatives, we find the nonlinear strain tensor used in the main text,
\begin{equation}
u_{ij}(\mathbf{x})=\frac{1}{2}\left(\partial_i u_j+\partial_j u_i +\partial_i f \partial_j f\right)-\delta_{ij}\frac{f}{R}.
\end{equation}
The stretching energy is then given by~\cite{si_landau_elasticity_1986}
\begin{equation}
G_s=\frac{1}{2}\int dS\,\left[2\mu u_{ij}^2 + \lambda u_{kk}^2\right],
\end{equation}
where $\mu$ and $\lambda$ are the  Lam\'e coefficients  and $dS$ is an area element.

We also include a bending energy of the Helfrich form~\cite{si_helfrich_elastic_1973} that penalizes changes in local curvature:
\begin{equation}
G_b = \frac{\kappa}{2}\int dS \, (H-H_0)^2,
\end{equation}
where $\kappa$ is the bending rigidity, $H$ the mean curvature and $H_0$ the spontaneous mean curvature (which we take to be equal everywhere to the curvature $2/R$ of the undeformed shell). For a shallow section of the shell, the local curvature can be written in terms of the height field $Z(x_1,x_2)+f(x_1,x_2)$ as 
\begin{equation}
H = \nabla^2(Z+f) = \frac{2}{R}+\nabla^2 f,
\end{equation}
where $\nabla^2 = \partial_{11}+\partial_{22}$ is the Laplacian in the tangential coordinate system. Finally the energy due to an external pressure $p$ equals the work done, 
\begin{equation}
W = -p\int dS \,f.
\end{equation}

The area element is $dS = dx_{1}dx_{2} /\sqrt{1-(x_{1}^{2}+x_{2}^{2})/R^{2}}\approx dx_1dx_2$ when terms of order $(x_{i}/R)^2$ and above are neglected. Summing the stretching, bending and pressure energies leads to the elastic energy expression $G=G_s+G_b+W$, Eq.~(1) in the main text. 

Since we are restricted to a shallow section of the shell, the theory is strictly applicable only to deformations whose length scale is small compared to the radius $R$. The typical length scale $\ell$ of deformations can be obtained by balancing the bending and stretching energies $G_b$ and $G_s$ discussed above. Upon noting that the stretching free energy density in a region of size $\ell$ is $\mathcal{G}_s \sim Y(f/R)^2$, where $Y$ is a typical elastic constant, and $\mathcal{G}_b \sim \kappa f^2/\ell^4$, we recover the F\"oppl-von K\'arm\'an length scale introduced in the main text, 
\begin{equation}
\ell^* = \frac{R}{\gamma^{1/4}},
\end{equation}
where the F\"oppl-von K\'arm\'an number is $\gamma = YR^2/\kappa$. More sophisticated calculations (sketched below) show that the relevant elastic constant is the 2D Young's modulus, $Y=4\mu(\mu+\lambda)/(2\mu+\lambda)$. 

For a shell made up of an elastic material of thickness $h$, taking $Y$ and $\kappa$ from the 3D Young's modulus of an isotropic solid within thin shell theory provides the estimate $\gamma \approx 10 (R/h)^2$~\cite{si_landau_elasticity_1986}. For shallow shell theory to be valid, we need $\ell^* \ll R$. Hence shallow shell theory is valid when $\gamma \gg 1$ i.e. $R \gg h$, which is precisely the limit of large, thin curved shells which are most susceptible to thermal fluctuations. This agreement between shallow shell theory and more general shell theories that are applicable over entire spherical shells has been discussed by Koiter~\cite{si_koiter_spherical_1963} in the context of the response of a shell to a point force at its poles. Shallow shell theory was also  used to study the stability of pressurized shells by Hutchinson~\cite{si_hutchinson_imperfection_1967}. In both cases, shallow shell theory was shown to be valid for thin shells such that $h/R \ll 1$. Since thermal fluctuations are only relevant for shells with radii several orders of magnitude larger than their thickness, shallow shell theory is an excellent starting point for the extremely thin shells of interest to us here.

\section*{Elimination of in-plane phonon modes and uniform spherical contraction by Gaussian integration}
A spherical shell under the action of a uniform external pressure that is lower than the critical buckling threshold responds by contracting uniformly by an amount $f_0$. The out-of-plane deformation field can then be written as a sum of its uniform and non-uniform parts, 
\begin{equation}
f(\mathbf{x}) = f_0 + f^\prime(\mathbf{x})= f_0 +\sum_{\mathbf{q} \neq 0 }f_\mathbf{q}e^{-i\mathbf{q}\cdot\mathbf{x}},
\end{equation}
where $f^\prime(\mathbf{x})$ represents the contribution to the field from its $\mathbf{q} \neq 0$ Fourier components. (In this section, for ease of presentation we use the normalization $f_\mathbf{q} \equiv \frac{1}{A}\int d^2x\,f(\mathbf{x})e^{i\mathbf{q}\cdot\mathbf{x}},$ where $A$ is the area of integration in the $(x_1,x_2)$ plane. The inverse transform is then $f(\mathbf{x})=\sum_{\mathbf{q}}f_\mathbf{q}e^{-i\mathbf{q}\cdot\mathbf{x}}.$) With this decomposition, $\int\,d^2x\,f^\prime(\mathbf{x})=0$ and thus only $f_0$ contributes to the pressure work $W$. On the other hand, only $f^\prime$ contributes to the nonlinear part of the strain tensor. Hence the elastic energy $G=G_b+G_s+W$ defined above is harmonic in the in-plane phonon fields $u_1(\mathbf{x})$ and $u_2(\mathbf{x})$ as well as the uniform contraction $f_0$. To analyze the effects of anharmonicity, it is useful to eliminate these fields and define an effective free energy~\cite{si_nelson_statistical_1988},
\begin{equation} \label{si_eqn_gaussianintegral}
\begin{split}
G&_\mathrm{eff}[f^\prime] = -k_\text{B}T \ln\left\{\int \mathcal{D}\vec{u}(x_1,x_2)\int df_0\,e^{-G[f^\prime ,f_0,u_1,u_2]/k_\text{B}T]}\right\}.
\end{split}
\end{equation}
To carry out the functional integrals in Eq.~(\ref{si_eqn_gaussianintegral}) for a fixed out-of-plane displacement field $f^\prime(\mathbf{x})$, the strain tensor $u_{ij}$ must also be separated into its $\mathbf{q}=0$ and $\mathbf{q}\neq 0$ components:
\begin{equation}
u_{ij}=\tilde{u}^0_{ij}+\sum_{\mathbf{q}\neq 0}\left[\frac{i}{2}\left(q_i u_j(\mathbf{q})+q_ju_i(\mathbf{q})\right)+A_{ij}(\mathbf{q})-\delta_{ij}\frac{f_\mathbf{q}}{R}\right]e^{-i \mathbf{q \cdot x}}
\end{equation}
where
\begin{equation}
A_{ij}(\mathbf{q})=\frac{1}{2A}\int d^2x\,\partial_i f^\prime\, \partial_j f^\prime\, e^{i \mathbf{q \cdot x}}.
\end{equation}
The uniform part of the strain tensor has the following components:
\begin{equation} \label{si_eqn_uij0}
\begin{split}
\tilde{u}^0_{11} &= u^0_{11} + A_{11}(\mathbf{0}) - \frac{f_0}{R}, \\
\tilde{u}^0_{22} &= u^0_{22} + A_{22}(\mathbf{0}) - \frac{f_0}{R},\\
\tilde{u}^0_{12} &= u^0_{12} + A_{12}(\mathbf{0}).\\
\end{split}
\end{equation}
Here, $u^0_{ij}$ are the uniform in-plane strains that are \emph{independent} of $f_0$. This restriction implies that $u^0_{11}+u^0_{22}=0$ because a simultaneous uniform in-plane strain of the same sign in the $x_1$ and $x_2$ directions corresponds to a change in radius of the sphere and thus cannot be decoupled from $f_0$. Hence in addition to $f_0$ and $u_{12}^0$, there is only one more independent degree of freedom, $\Delta u^0\equiv u_{11}^0-u_{22}^0$, that determines the uniform contribution to the strain tensor.

Finally we perform the functional integration in Eq.~(\ref{si_eqn_gaussianintegral}) over the phonon fields $u_i$ as well as the three independent contributions to the uniform part of the strain tensor --- $f_0$, $\tilde{u}_{12}^0$ and $\Delta u^0$. The resulting effective free energy is, upon suppressing an additive constant,

\begin{equation} \label{si_eqn_feff}
G_\mathrm{eff} = \int d^2 x\left[\frac{\kappa}{2}(\nabla^2 f^\prime)^2+\frac{Y}{2} \left(\frac{1}{2}P^\mathrm{T}_{ij}\partial_i f^\prime \partial_j f^\prime-\frac{f^\prime}{R}\right)^2\right]-A\frac{pR}{2}\left[A_{11}(\mathbf{0})+A_{22}(\mathbf{0})\right]
\end{equation}

where $P^\mathrm{T}_{ij} = \delta_{ij}-\partial_i \partial_j/\nabla^2$ is the transverse projection operator. Note that as a result of the integration the Lam\'e coefficients $\mu$ and $\lambda$ enter only through the 2D Young's modulus $Y = 4\mu(\mu+\lambda)/(2\mu+\lambda)$. Finally, substituting 

\begin{equation}
A_{11}(\mathbf{0})+A_{22}(\mathbf{0}) = \frac{1}{2A}\int d^2x\,\left[(\partial_1 f^\prime)^2+(\partial_2 f^\prime)^2\right] = \frac{1}{2A}\int d^2x|\nabla f^\prime|^2
\end{equation}

in Eq.~(\ref{si_eqn_feff}) gives the effective free energy used in the analysis, Eq.~(3) in the main text. In the following, we drop the prime on the out-of-plane displacement field since $f_0$ has now been eliminated. When only the harmonic contributions are considered, the equipartition result for the thermally generated Fourier components $f_\mathbf{q} = \int d^2x \,f(\mathbf{x})\exp(i\mathbf{q}\cdot\mathbf{x})$ with two-dimensional wavevector $\mathbf{q}$ are
\begin{equation} \label{si_eqn_corrfn_gaussian}
\langle f_\mathbf{q}f_\mathbf{q^\prime} \rangle_0 = \frac{Ak_\mathrm{B}T \delta_{\mathbf{q},\mathbf{-q^\prime}}}{\kappa q^4 -\frac{pR}{2}q^2+ \frac{Y}{R^2}}.
\end{equation}
where $A$ is the area of integration in the $(x_1,x_2)$ plane. This harmonic spectrum [Eq.~(4) in the main text] takes on corrections due to the anharmonic terms that are calculated in the next section.

\section*{One-loop contributions to the self-energy}

\begin{figure*}[t] 
\centering
\includegraphics[width=7in]{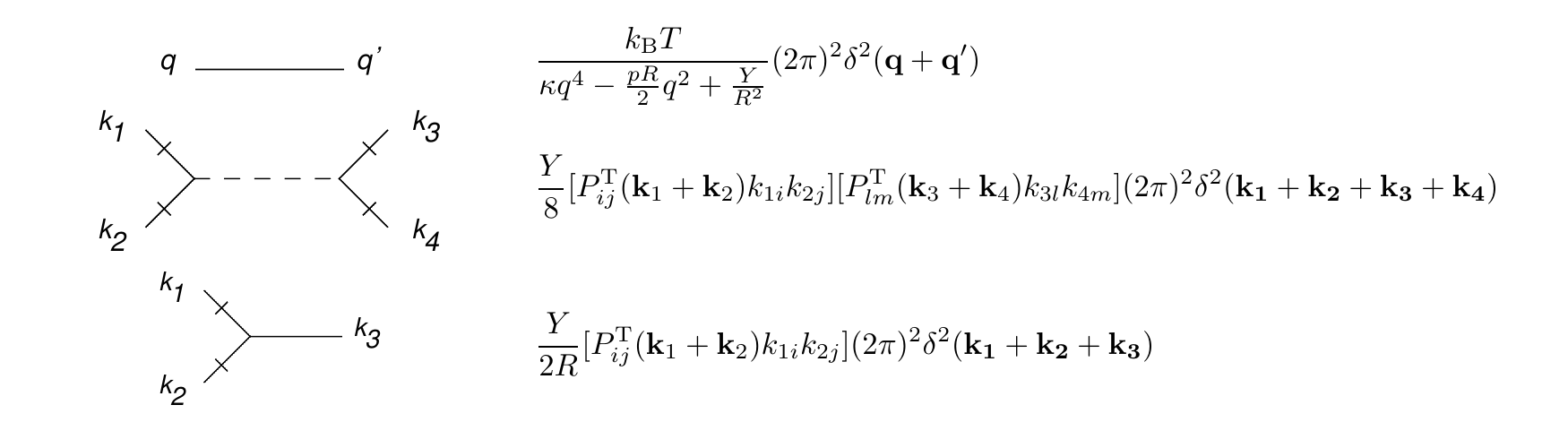}
\caption{The bare propagator for $f(\mathrm{q})$ and the vertices arising from the non-quadratic terms in $G_\mathrm{eff}$. The slashes on specific legs denote spatial derivatives.  $P^\mathrm{T}_{ij}(\mathbf{q}) = \delta_{ij}-q_iq_j/q^2$ is the transverse projection operator in momentum space. Note an unusual feature of this graphical perturbation theory: the system size, \emph{i.e.} the sphere radius $R$, enters explicitly both in the propagator and as a coupling constant in the third order interaction vertex.} \label{si_tab_feynmanrules}
\end{figure*}

\begin{figure*}[t] 
\includegraphics[width=3.42in]{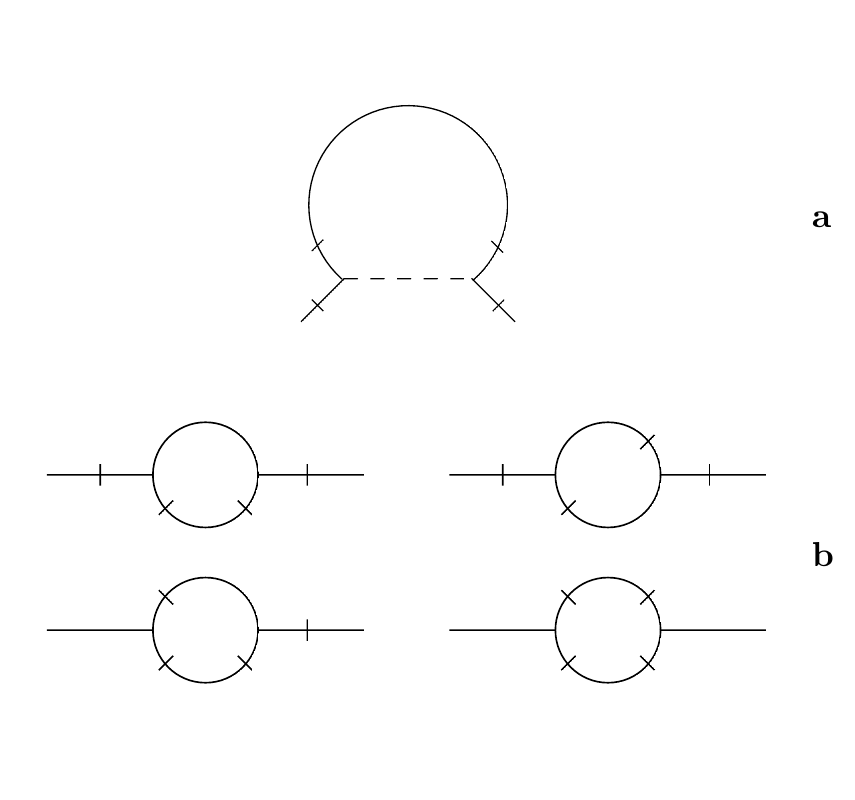}
\caption{One-loop corrections to the two-point height-height correlation function in momentum space. Note that in calculating the self-energy, the external propagators are not included; \emph{i.e.} they are ``amputated''. The contribution in {\bf a} resembles that for membranes with a flat ground state, except for the $R$-dependent pressure and mass terms in the propagator. The nonlinear corrections in {\bf b}, however, arise from a cubic coupling  constant proportional to $1/R$ and are unique to the spherical geometry.} \label{si_fig_oneloop}
\end{figure*}

Here we describe the self-energy used to calculate the leading anharmonic corrections to the fluctuation spectrum in the main text. The Feynman rules obtained from the effective free energy $G_\mathrm{eff}[f]$ are summarized in Fig.~S2. Henceforth, Fourier components are defined as in the main text: $f_\mathbf{q} = \int d^2x \,f(\mathbf{x})\exp(i\mathbf{q}\cdot\mathbf{x})$ with two-dimensional wavevector $\mathbf{q}$. The inverse Fourier transformation of the out-of-plane deformation field is 
\begin{equation}
    f(\mathbf{x}) = \frac{1}{A}\sum_{\mathbf{q}\neq\mathbf{0}}f_{\mathbf{q}}e^{-i\mathbf{q}\cdot\mathbf{x}},
\end{equation}
where $A$ is the area of integration in the $(x_1,x_2)$ plane and the sum is over all allowed Fourier modes. The one-loop contribution to the self-energy $\Sigma(\mathbf{q})$ due to the anharmonic three-point vertex (cubic term in the energy) and the four-point vertex (quartic term) are summarized in Fig.~S3. Fig.~S3a is also present in the calculation for flat membranes~\cite{si_nelson_statistical_1988}, and provides a contribution
\begin{equation} \label{si_eqn_hat}
-Y\int \frac{d^2 k}{(2\pi )^2}\,\frac{[P^\mathrm{T}_{ij}(\mathbf{k})q_i q_j]^2}{\kappa|\mathbf{q+k}|^4 -\frac{pR}{2}|\mathbf{q+k}|^2+\frac{Y}{R^2}}
\end{equation}
to the self-energy. Fig.~S3b involves two-vertex terms arising from the cubic coupling unique to shells with curvature (note that, despite ``amputation'' of the propagator legs, the diagrams are distinct because the slashes decide the momentum terms which survive various index contractions in addition to determining the momentum of the transverse projection operator introduced at each vertex). The net contribution to the self-energy from the four diagrams in Fig.~S3b is

\begin{equation} \label{si_eqn_candy}
\begin{split}
&\frac{Y^2}{R^2}\int\frac{d^2 k}{(2\pi )^2}\,\frac{1}{\Bigl(\kappa|\mathbf{q+k}|^4-\frac{pR}{2}|\mathbf{q+k}|^2+\frac{Y}{R^2}\Bigr)\Bigl(\kappa k^4-\frac{pR}{2}q^2+\frac{Y}{R^2}\Bigr)}\times \\ &\quad\left\{\frac{1}{2}[P^\mathrm{T}_{ij}(\mathbf{q})k_i k_j]^2+[P^\mathrm{T}_{ij}(\mathbf{k})q_i q_j]^2+[P^\mathrm{T}_{ij}(\mathbf{k})q_i q_j][P^\mathrm{T}_{lm}(\mathbf{k+q})q_l q_m]+2[P^\mathrm{T}_{ij}(\mathbf{k})q_i q_j][P^\mathrm{T}_{lm}(\mathbf{q})k_l k_m]\right\}
\end{split}
\end{equation}

While the inverse of the harmonic correlation function, Eq.~(\ref{si_eqn_corrfn_gaussian}), only contains terms of order $q^{0}$, $q^{2}$, $q^{4}$, the one-loop corrections to the spectrum [Eqs.~(\ref{si_eqn_hat}--\ref{si_eqn_candy})] generate terms with these powers of $q$ as well as terms of order $q^{6}$ and above in the full inverse fluctuation spectrum. If we keep only terms of order $q^{4}$ and below in the calculation of the one-loop fluctuation spectrum, we can provide an approximate description of the low-$q$ behaviour of the shell in terms of effective elastic constants:
\begin{equation}
Ak_\mathrm{B}T\langle|f_\mathbf{q\rightarrow 0}|^2\rangle^{-1} \equiv \kappa_{\scriptscriptstyle\mathrm{R}} q^4 - \frac{p_{\scriptscriptstyle\mathrm{R}}R}{2} q^2+\frac{Y_{\scriptscriptstyle\mathrm{R}}}{R^2} + O(q^6),
\end{equation}
where $Y_{\scriptscriptstyle\mathrm{R}}$, $\kappa_{\scriptscriptstyle\mathrm{R}}$ and $p_{\scriptscriptstyle\mathrm{R}}$ are the effective Young's modulus, bending rigidity and dimensionless pressure respectively. At long length scales, probes of the elastic properties of thermally fluctating shells would provide information of these effective elastic constants rather than the ``bare'' constants $Y$, $\kappa$ and $p$ that describe the zero-temperature shell. Upon expanding the integrands in Eqs.~(\ref{si_eqn_hat}--\ref{si_eqn_candy}) to O($q^{4}$) the momentum integrals can be carried out analytically to obtain:

\begin{equation} \label{si_eqn_yr_whole}
Y_{\scriptscriptstyle\mathrm{R}} = Y\left[1-\frac{3}{128\pi}\frac{k_\text{B}T}{\kappa}\frac{\sqrt\gamma}{(1-\eta^2)^{3/2}}\left(\eta\sqrt{1-\eta^2}+\pi - \cos^{-1}\eta\right)\right],
\end{equation}
\begin{equation} \label{si_eqn_kr_whole}
\begin{split}
\kappa_{\scriptscriptstyle\mathrm{R}} &= \kappa\Biggl[1+\frac{1}{30720 \pi}\frac{k_\text{B}T}{\kappa}\frac{\sqrt\gamma}{(1-\eta^2)^{7/2}}\biggl[\eta\sqrt{1-\eta^2}\left(-1699+3758\eta^2-2104\eta^4\right) \\ &\qquad\qquad\qquad
+15(61-288\eta^2+416\eta^4-192\eta^6)\left(\pi - \cos^{-1}\eta\right)\biggr]\Biggr],
\end{split}
\end{equation}
\begin{equation} \label{si_eqn_pr_whole}
\eta_{\scriptscriptstyle\mathrm{R}}=\eta+\frac{1}{1536 \pi}\frac{k_{B}T}{\kappa} \frac{\sqrt{\gamma }}{\left(1-\eta ^2\right)^{5/2}} \left[\sqrt{1-\eta ^2} \left(64-67 \eta
   ^2\right)+3\left(21 \eta -22 \eta ^3\right)
   \left(\pi-\cos ^{-1}\eta\right)\right].
\end{equation}

where we have defined a dimensionless pressure $\eta \equiv p/p_{\text{c}}$ and $p_{\text{c}} = 4\sqrt{\kappa Y}/R^{2}$ is the classical buckling pressure of the shell. We see explicitly that the quantities diverge in the limit $\eta \to 1$. To lowest order in the external pressure, we have
\begin{equation} \label{si_eqn_smallpyr}
Y_{\scriptscriptstyle\mathrm{R}} \approx  Y \left[1  -\frac{3}{256}\frac{k_\mathrm{B}T}{\kappa}\sqrt{\gamma}\left(1+\frac{4}{\pi}\frac{p}{p_{\text{c}}}\right)\right],
\end{equation}
\begin{equation} \label{si_eqn_genpressure}
p_{\scriptscriptstyle\mathrm{R}} \approx p +\frac{1}{24\pi}\frac{k_\mathrm{B}T}{\kappa}p_\text{c}\sqrt{\gamma}\left(1+\frac{63\pi}{128}\frac{p}{p_{\text{c}}}\right) ,
\end{equation}
and
\begin{equation}\label{si_eqn_smallpkr}
\kappa_{\scriptscriptstyle\mathrm{R}} \approx \kappa\left[1 +\frac{61}{4096}\frac{k_\mathrm{B}T}{\kappa}\sqrt{\gamma}\left(1-\frac{1568}{915\pi}\frac{p}{p_{\text{c}}}\right)\right].
\end{equation}
These are the approximate renormalized elastic quantities tabulated in Eqs.~(7--9) in the main text.

In evaluating the above expressions, the momentum integrals in Eqs.~(\ref{si_eqn_hat}--\ref{si_eqn_candy}) must strictly speaking be carried out over the phase space of all allowed Fourier modes $f(\mathbf{k})$ of the system, which go from some low-$k$ cutoff $k_{\text{min}}\sim 1/R$ to a high-$k$ cutoff set by the microscopic lattice constant. However, since all integrals converge in the ultraviolet limit $k \to \infty$, the upper limit of the $k$-integrals can be extended to $\infty$. The integrals are well-behaved at low momenta due to the mass term $\sim Y/R^2$ in the propagator. Hence we carry out the momentum integrals over the entire two-dimensional plane of $\mathbf{k}$. The excess contribution to the self energy by including spurious Fourier modes with $0< k < 1/R$, \emph{i.e.} for wavevectors less than the natural infrared cutoff $k_\mathrm{min} \sim 1/R$,  gives rise to an error of roughly $1/\sqrt{\gamma}$ which is negligible for extremely thin shells. This correction is of similar magnitude to the errors introduced by using shallow shell theory (which is inaccurate for the longest-wavelength modes with wavevector $k\sim 1/R$) which are also negligible in the thin-shell limit.

\section*{Calculation of fluctuation spectrum with spherical harmonics}
While the perturbation theory calculations were carried out using a basis of Fourier modes in a shallow section of the shell to decompose the radial displacement field, the fluctuation spectrum is most efficiently measured in simulations using a spherical harmonics expansion. To compare the simulation results to the expected corrections from perturbation theory, we use the description of the shell in terms of the effective elastic constants $Y_{\scriptscriptstyle\mathrm{R}}$, $\kappa_{\scriptscriptstyle\mathrm{R}}$ and $p_{\scriptscriptstyle\mathrm{R}}$, Eqs.~(7--9) in the main text. 

Consider a spherical shell of radius $R$ with bending rigidity $\kappa$ and Lam\'e coefficients $\lambda$ and $\mu$, experiencing a tangential displacement field $\mathbf{u} = (u_{x},u_{y})$ and a radial displacement field $f$. Like any smooth vector field, $\mathbf{u}$ can be decomposed into an irrotational (curl-free) part and a solenoidal (divergence-free) part: $\mathbf{u} \equiv \nabla\Psi+\mathbf{v}$, where the scalar function $\Psi$ generates the irrotational component and $\mathbf{v}$ is the solenoidal component. Upon expanding $f\equiv \sum_{l,m}A_{lm}RY_{l}^{m}$ and $\Psi \equiv \sum_{l,m}B_{lm}R^{2}Y_{l}^{m}$ in terms of spherical harmonics $Y_{l}^{m}(\theta,\phi)$, the elastic energy of the deformation to quadratic order in the fields is given by~\cite{si_zhang_scaling_1993}
\begin{equation}
\begin{split}
    G &= R^{2}\sum_{l,m}\biggl\{\left[\frac{\kappa}{2}\frac{(l+2)^{2}(l-1)^{2}}{R^{2}}+2K\right] A_{lm}^{2}-2Kl(l+1)A_{lm}B_{lm} \\
    &\qquad\qquad\qquad+\frac{1}{2}l(l+1)\left[(K+\mu)l(l+1)-2\mu\right]B_{lm}^{2}\biggr\}+G_{\text{sol}}(\mathbf{v}),
\end{split}
\end{equation}
where $K=\lambda+\mu$ is the bulk modulus. The solenoidal component $\mathbf{v}$ does not couple to the radial displacement field and provides an independent contribution $G_{\text{sol}}$ which is purely quadratic in the field $\mathbf{v}$. 

To this elastic energy, we also add the surface energy-like contribution $G_{\text{S}}=-(pR/2) \Delta A$ due to the ``negative surface tension'' $-pR/2$  present in the shell when it is uniformly compressed in response to an external pressure $p$. Here $\Delta A$ is the excess area due to deformations about the average radius. In terms of spherical harmonic coefficients, this area change can be written~\cite{si_milner_dynamical_1987}
\begin{equation}
    \Delta A \approx R^{2}\sum_{l>1,m}A_{lm}^{2}\left[1+\frac{l(l+1)}{2}\right].
\end{equation}

As we did for the elastic energy in shallow shell theory, we can now integrate out the quadratic fluctuating quantities $B_{lm}$ and the solenoidal field $\mathbf{v}$ to obtain an effective free energy in terms of the radial displacements alone:
\begin{equation}
    G_{\text{eff}} = \frac{R^{2}}{2}\sum_{l>1,m}\left\{\frac{\kappa(l+2)^{2}(l-1)^{2}}{R^{2}}-pR\left[1+\frac{l(l+1)}{2}\right]+\frac{4\mu(\mu+\lambda)(l^{2}+l-2)}{(2\mu+\lambda)(l^{2}+l)-2\mu}\right\}A_{lm}^{2}.
\end{equation}
The fluctuation amplitude is obtained via the equipartition theorem:
\begin{equation}
    \begin{split}
        k_{B}T\langle |A_{lm}|^{2}\rangle_{0}^{-1} &= \kappa(l+2)^{2}(l-1)^{2}-pR^{3}\left[1+\frac{l(l+1)}{2}\right]+\frac{4\mu(\mu+\lambda)(l^{2}+l+2)}{(2\mu+\lambda)(l^{2}+l)-2\mu}R^{2} \\
    &= \kappa(l+2)^{2}(l-1)^{2}-pR^{3}\left[1+\frac{l(l+1)}{2}\right]+\frac{Y}{1+\frac{Y}{2\mu(l^{2}+l-2)}}R^{2}. \label{si_eqn.sphfluct}
    \end{split}
\end{equation}
where $Y = 4\mu(\mu+\lambda)/(2\mu+\lambda)$ is the 2D Young's modulus introduced earlier. The effect of anharmonic contributions to the fluctuation spectrum can now be calculated by using the effective temperature-dependent quantities $Y_{\scriptscriptstyle\mathrm{R}}$, $\kappa_{\scriptscriptstyle\mathrm{R}}$ and $p_{\scriptscriptstyle\mathrm{R}}$ in place of the bare elastic constants in the above expression. However, the last term in Eq.~(\ref{si_eqn.sphfluct}) also requires knowledge of the thermal corrections to the Lam\'e coefficient $\mu$ which was eliminated in the shallow shell calculation when the tangential displacement fields were integrated out. For the discretized stretching energy used in the simulations, we have $\mu = 3Y/8$. If we assume that this relationship is not significantly changed by the anharmonic corrections to one-loop order, then $\mu_{\scriptscriptstyle\mathrm{R}} \approx 3Y_{\scriptscriptstyle\mathrm{R}}/8$. Upon substituting this approximation together  with the other effective elastic parameters in Eq.~(\ref{si_eqn.sphfluct}), we find
\begin{equation}
    \begin{split}
        k_{B}T\langle |A_{lm}|^{2}\rangle^{-1} \approx &\kappa_{\scriptscriptstyle\mathrm{R}}(l+2)^{2}(l-1)^{2} -p_{\scriptscriptstyle\mathrm{R}}R^{3}\left[1+\frac{l(l+1)}{2}\right]  +Y_{\scriptscriptstyle\mathrm{R}}R^{2}\left[\frac{3(l^{2}+l-2)}{3(l^{2}+l)-2}\right] \label{si_eqn_approxsph}
    \end{split}
\end{equation}
which is the same as as Eq.~(11) in the main text~\footnote{If, as is more likely, the thermal corrections to $\mu$ and $Y$ do differ to O($k_{B}T$), we can nevertheless estimate that the resulting error term introduced by the assumption $\mu_{\scriptscriptstyle\mathrm{R}} \approx 3Y_{\scriptscriptstyle\mathrm{R}}/8$ is suppressed by a factor $4/[3(l^{2}+l-2)+4]$ relative to the anharmonic corrections and is thus atleast an order of magnitude smaller than the anharmonic contribution itself when $l>1$.}. 

\section*{Linear response of the shell to point forces}
We calculate the response of the shallow shell to a point force at the origin, corresponding to a force field $h(\mathbf{x}) = F\delta^{2}(\mathbf{x})$. The Fourier decomposition of this force field is
\begin{equation}
    h_\mathbf{q} = F, \,\text{for all}\, \mathbf{q}.
\end{equation}
The linear response of the deformation field $f$ to this force is related to its fluctuation amplitudes in the \emph{absence} of the force, $\langle |f_\mathbf{q}|^2\rangle_{h=0}$, by the fluctuation-response theorem:
\begin{equation}
    \langle f_\mathbf{q} \rangle  = \frac{\langle |f_\mathbf{q}|^2\rangle_{h=0}}{Ak_\text{B}T}h_{\mathbf{q}} = \frac{\langle |f_\mathbf{q}|^2\rangle_{h=0}}{Ak_\text{B}T}F.
\end{equation}
The inward deflection at the origin is then
\begin{equation} \label{si_eqn0deflection}
    \langle f(\mathbf{x}=0)\rangle = \frac{1}{A} \sum_{\mathbf{q}}\langle f_\mathbf{q} \rangle = \frac{F}{A^{2}k_{B}T}\sum_{\mathbf{q}} \langle |f_\mathbf{q}|^2\rangle_{h=0}.
\end{equation}
This can be related to $\langle f^{2} \rangle$, the mean square fluctuations of the deformation field in real space which is a position-independent quantity in the absence of nonuniform external forces:
\begin{equation} \label{si_eqn0msfluct}
\begin{split}
    \langle f^{2} \rangle \equiv \langle [f(\mathbf{x})]^{2} \rangle_{h=0} &= \frac{1}{A^{2}}\sum_{\mathbf{q}}\sum_{\mathbf{q^{\prime}}}\langle f_{\mathbf{q}}f_{\mathbf{q^{\prime}}}\rangle e^{-i(\mathbf{q}+\mathbf{q^{\prime}})\cdot\mathbf{x}} \\
    &= \frac{1}{A^{2}}\sum_{\mathbf{q}}\langle |f_\mathbf{q}|^2\rangle_{h=0}.
\end{split}
\end{equation}
From Eqs.~(\ref{si_eqn0deflection}) and (\ref{si_eqn0msfluct}), we obtain
\begin{equation} \label{si_eqn0defl-fluct}
    \langle f(\mathbf{x}=0)\rangle = \frac{F}{k_{B}T}\langle f^{2}\rangle.
\end{equation}
This equation relates the depth of the indentation due to a force $F$ at the origin to the mean square fluctuations of the deformation field $f$ in the absence of such a force.
    
When only harmonic contributions are considered, Eq.~(\ref{si_eqn_corrfn_gaussian}) gives us the mean square amplitude  $\langle |f_\mathbf{q}|^2\rangle_0 = Ak_{B}T/(\kappa q^4 - pRq^2/2+Y/R^{2})$ in terms of the elastic constants and external pressure. Upon taking the continuum limit of the sum over wavevectors $\sum_{\mathbf{q}} \to A\int d^{2}q/(2\pi)^{2}$, we can calculate the fluctuation amplitudes exactly:
\begin{equation} \label{si_eqn0bareflucts}
    \langle f^{2} \rangle = \int\frac{d^{2}q}{(2\pi)^{2}}\frac{k_{B}T}{\kappa q^4-\frac{pR}{2}q^{2}+\frac{Y}{R^{2}}} = \frac{Rk_{B}T}{8\sqrt{\kappa Y}}\frac{1+\frac{2}{\pi}\sin^{-1}\eta}{\sqrt{1-\eta^{2}}},
\end{equation}
where $\eta = p/p_{\text{c}}= pR^{2}/(4\sqrt{\kappa Y})$ is the dimensionless pressure, and $\eta < 1$, \emph{i.e.} we restrict ourselves to pressures below the classical buckling pressure. From Eqs.~(\ref{si_eqn0defl-fluct}) and (\ref{si_eqn0bareflucts}), we get the linear relation between the indentation force and the depth of the resulting deformation:
\begin{equation}
    F=\frac{8\sqrt{\kappa Y}}{R}\frac{\sqrt{1-\eta^{2}}}{1+\frac{2}{\pi}\sin^{-1}\eta}\langle f(\mathbf{x}=0)\rangle.
\end{equation}
The temperature drops out and we obtain a result valid for $T=0$ shells as well. The expression reproduces the well-known Reissner solution~\cite{si_reissner_stresses_1946} for the linear response of a spherical shell to a point force when $\eta=0$, and also reproduces the recent result from Vella et al~\cite{si_vella_indentation_2011} for indentations on spherical shells with an \emph{internal} pressure when $\eta < 1$. At finite temperatures, however, anharmonic effects contribute terms of order $(k_{B}T)^{2}$ and higher to $\langle f^{2} \rangle$, making the response temperature-dependent.

In the simulations, the shells contract by a small amount due to thermal fluctuations, even in the absence of external forces. Thus, indentations are measured relative to the thermally averaged pole-to-pole distance of the shell at finite temperature, $\langle z_0 \rangle < 2R$. Equal and opposite inward forces are applied to the north and south poles of the shell to maintain a force balance (see details in the {\it Materials and Methods} section of the main text) and the resulting average pole-to-pole distance, $\langle z \rangle$, is measured. This corresponds to an average indentation depth of $(\langle z_{0} \rangle - \langle z \rangle)/2$ at each pole, with associated force [from Eq.~(\ref{si_eqn0defl-fluct})]
\begin{equation}
    F = \frac{k_{B}T}{\langle f^{2} \rangle}\frac{(\langle z_{0} \rangle - \langle z \rangle)}{2} \equiv k_{\text{s}} (\langle z_{0} \rangle - \langle z \rangle),
\end{equation}
\emph{i.e.} the shell as a whole acts as a spring with spring constant 
\begin{equation} \label{si_eqn0springconstant}
    k_{\text{s}} = \frac{k_{B}T}{2\langle f^{2} \rangle}.
\end{equation}

At $T=0$, we have 
\begin{equation} \label{si_eqn0springconstantt0}
    k_{\text{s}} = \frac{4\sqrt{\kappa Y}}{R}\frac{\sqrt{1-\eta^{2}}}{1+\frac{2}{\pi}\sin^{-1}\eta};
\end{equation}
in particular, $k_{\text{s}}=4\sqrt{\kappa Y}/R$ in the absence of external pressure. Anharmonic contributions change the fluctuation amplitude $\langle f^{2} \rangle$ and hence the linear response. To lowest order in temperature, the effects of anharmonic contributions can be obtained by using the renormalized elastic constants calculated using perturbation theory [Eqs.~(\ref{si_eqn_smallpyr})--(\ref{si_eqn_smallpkr})] in Eq.~(\ref{si_eqn0springconstantt0}) and keeping terms to $O(T)$. In particular, even if the bare pressure $p = 0$, the renormalized dimensionless pressure $p_{\text{R}}$ is nonzero and affects the spring constant, as do the temperature-dependent effective elastic moduli. The result in this case is
\begin{equation}
    k_{\text{s}}(T>0) \approx \frac{4\sqrt{\kappa Y}}{R}\left[1-0.0069\frac{k_{B}T}{\kappa} \sqrt\gamma\right].
\end{equation}
This is the theoretical prediction quoted as Eq.~(13) in the main text.

\section*{Measuring the effective spring constant from Monte Carlo simulations}

\begin{figure*}[t]
\centering
\includegraphics[width=7in]{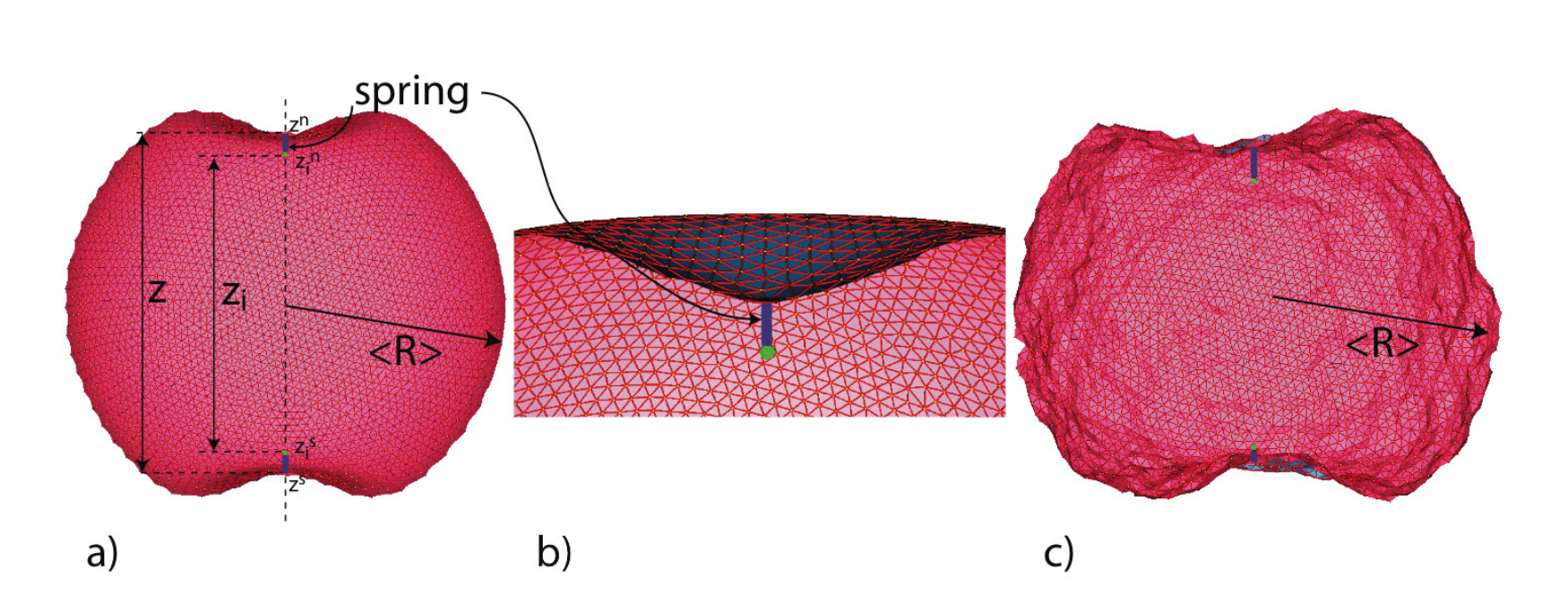}
\caption{Illustration of the indentation simulation of a randomly triangulated shell. Shown here are vertical cuts through a shell of radius $R_0=20 \ r_0$ at dimensionless temperature $k_BT/\kappa\sqrt{\gamma}=10^{-5}$ ({\bf a} \& {\bf b}) and  $k_BT/\kappa\sqrt{\gamma}=15$ ({\bf c}).  A shell at low temperature ({\bf a}) is indented by two harmonic springs (dark blue lines) attached  to the north ($z^\text{N}$) and south ($z^\text{S}$) pole of the shell. Fixing the springs at a separation $z_i=z_\text{i}^\text{N}-z_\text{i}^\text{S}$ leads to a pole separation $z$. The green points indicate the positions $z_\text{i}^\text{N}$ and $z_\text{i}^\text{S}$ of the fixed end points of the springs. ({\bf b}) A close-up of the north pole of the shell displayed in {\bf a}. The configuration contains a minority of 5 and 7-fold coordinated vertices in addition to 6-fold coordinated ones. ({\bf c}) Illustration of a fluctuating shell at $R_0=20 \ r_0$ and $k_BT/\kappa\sqrt{\gamma}=10^{-5}$.} \label{si_fig_indent}
\end{figure*}

We extract the spring constants of thermally fluctuating shells for Fig. 3c in the main text by using the relation between $k_{\text{s}}$ and fluctuations in the transverse displacement field $f$ [Eq.~(\ref{si_eqn0springconstant})]. It is straightforward to measure the average pole-to-pole distance of the fluctuating shell in the {\it absence} of external forces, $\langle z_{0} \rangle = \langle R - f_{\text{N}} - f_{\text{S}} \rangle$, where $f_{\text{N}}$ and $f_{\text{S}}$ are the inward displacements at the north and south poles respectively. Since the displacements at the poles are expected to be independent of each other, the mean squared fluctuations in $z_{0}$ are closely related to the mean square fluctuations in $f$:
\begin{equation}
\langle z_{0}^{2}\rangle - \langle z_{0}\rangle^{2} \approx 2\langle f^2 \rangle.{}
\end{equation}
The spring constant can thus be measured indirectly from the fluctuations in the pole-to-pole distance using Eq.~(\ref{si_eqn0springconstant}):
\begin{equation}
    k_{\text{s}} = \frac{k_{B}T}{2\langle f^{2} \rangle} \approx \frac{k_{B}T}{\langle z_{0}^{2}\rangle - \langle z_{0}\rangle^{2}}.
\end{equation}

\end{document}